\documentclass[twocolumn,showpacs,preprintnumbers,fleqn,amsmath,amssymb,aps]{revtex4}
\usepackage{graphicx}
\usepackage{bm}
\usepackage{amsmath}

\begin{document}
\title{Efficient current-induced domain-wall displacement in ${\rm SrRuO_3}$}

\author{ Michael Feigenson$^1$}
\author{James W. Reiner$^2$}
\author{Lior Klein$^1$}
\affiliation{$^1$Department of Physics, Bar-Ilan University, Ramat-Gan 52900, Israel \\
$^2$Department of Applied Physics, Yale University, New Haven, Connecticut 06520-8284}


\date{\today}


\begin{abstract}

We demonstrate current-induced displacement of ferromagnetic domain
walls in sub-micrometer fabricated patterns of ${\rm SrRuO_3}$
films. The displacement, monitored by measuring the extraordinary
Hall effect, is induced at zero applied magnetic field and its
direction is reversed when the current is reversed.  We find that
current density in the range of $10^9 - 10^{10} \ {\rm A/m^2}$ is
sufficient for domain-wall displacement when the depinning field
varies between 50 to 500 Oe. These results indicate relatively high
efficiency of the current in displacing domain walls which we
believe is related to the narrow width ($\sim3$ nm) of domain walls
in this compound.

\end{abstract}

\maketitle

Occasionally, the functionality of spintronic devices requires the
ability to control the magnetic configuration of sub-micrometer
magnetic structures. Currently, this goal is achieved in many cases
(for instance in novel MRAM devices) by current lines external to
the magnetic region that generate  Lorentz forces; however, this
method suffers from non-locality (due to the weak spatial decay of
the magnetic field) and lack of scalability (since the needed
current in these lines does not decrease with the magnetic bit
size). Consequently, there is an intensive effort to find methods of
manipulating the magnetic configuration of such structures by
injecting spin-polarized electric current into the target region, a
process both local and scalable. In addition to their enormous
technological importance, such methods involve fascinating
theoretical issues; hence, intensive theoretical and experimental
efforts are invested in their study.

Two main relevant effects are considered in this respect: (a) the
interaction between spin-polarized current injected into a small
magnetic region and the magnetic moments of that region, and (b) the
interaction between spin polarized current and a ferromagnetic
domain wall. The first effect, suggested by Slonzcewski \cite{slon},
yields magnetic switching while the second effect, suggested  by
Berger \cite{berger},  yields domain wall displacement. Here we
address the latter effect.

\begin{figure}
\begin{center}
\includegraphics [scale=0.6]{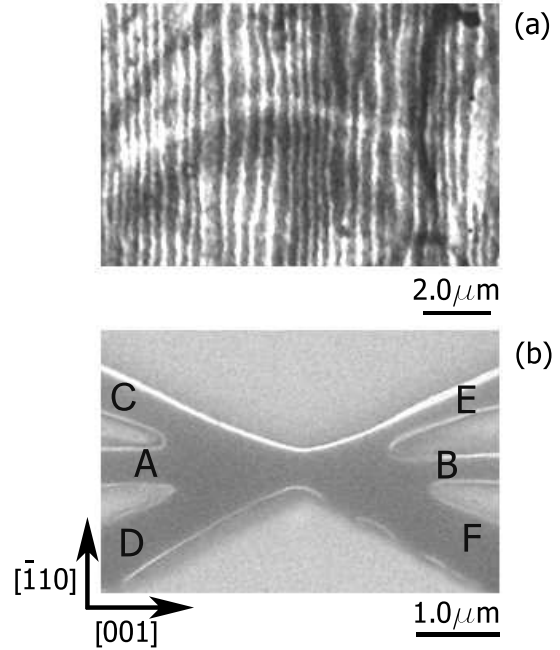}
\end{center}
\caption{(a) Image of magnetic domains in ${\rm SrRuO_3}$ with
transmission electron microscope in Lorentz mode (taken from
\cite{ann}). Bright and dark lines image domain walls at which
electron beam converges or diverges, respectively. Background
features are related to buckling of the film and are not related to
magnetic variations. (b) SEM image of the patterned sample. Current
pulses are injected between A and B (perpendicular to the domain
walls) and the average magnetization is sensed by measuring the
extraordinary Hall effect between C-D and between E-F. }
\label{pattern}
\end{figure}

Following the theoretical prediction of Berger \cite{berger}, the
interaction of spin-polarized current with a ferromagnetic wall has
been studied both theoretically and experimentally. There have been
various demonstrations of current-driven domain wall displacement
\cite{tsoi,groll,klaui,yama,lim,vernier,yama2,lauf} using various
methods for detecting the displacement, including magnetic force
microscopy \cite{tsoi,yama}, change in resistance of spin valve
structures \cite{groll,lim}, change in longitudinal resistance
\cite{klaui,lauf}, magneto-optics \cite{vernier,yama2} and Hall
effect \cite{yama2}. While current-driven domain wall motion in
ferromagnetic semiconductor systems is achieved with moderate
current density of less than $10^{9} {\rm  A/m^2}$ \cite{yama2}, the
typical current density for inducing domain-wall displacement in
metallic systems is in the range of $10^{11}-10^{12} {\rm  A/m^2}$.
Domain wall displacement at lower currents densities (on the order
of $10^{10} {\rm  A/m^2}$) in metallic systems is achieved either
when the depinning fields are very week (few Oersteds) or when the
domain wall resonance in its pinning potential is used
\cite{saltoh,luc}.

It has been suggested \cite{tatara} that two main mechanisms are
responsible for current-driven domain wall displacement: spin
transfer, expected to be relevant for wide walls, and momentum
transfer, expected to be relevant for narrow walls. The previously
studied metallic systems are all in the wide wall limit. Here we
study current-driven domain walls in  ${\rm SrRuO_3}$ which is an
excellent example of the narrow-wall limit.

We find that in  ${\rm SrRuO_3}$ depinning current ($J_c$) densities
in the range of $5.3\times 10^9 - 5.8\times 10^{10} \ {\rm A/m^2}$
induce domain wall displacement where the corresponding  depinning
fields ($H_c$) are between 50 to 500 Oe. For  comparison between
current-induced domain-wall displacement in different systems we
define $H_c/J_c$ as a measure of efficiency and find that the
efficiency in ${\rm SrRuO_3}$ is more than an order of magnitude
higher than in previously studied metallic systems of wide
domain-wall ferromagnets. While we believe that the high efficiency
is related to the fact that ${\rm SrRuO_3}$ is in the narrow
domain-wall limit, we find only partial agreement with relevant
theoretical predictions \cite{tatara}.

${\rm SrRuO_3}$ is a metallic perovskite with orthorhombic structure
(a=5.53, b=5.57, c=7.82 {\AA}) and an itinerant ferromagnet with
Curie temperature (for films)   of $ {\rm \sim 150 \ K }$. Our
samples are high-quality epitaxial thin films of ${\rm SrRuO_3}$
grown by reactive electron beam coevaporation on slightly miscut
$(\sim 0.2^o)$ ${\rm SrTiO_3}$ substrates with the [001] and $[{\bar
1}10]$ axes in the film plane. These films exhibit large uniaxial
magnetocrystalline anisotropy (anisotropy field of $\sim 10$ T) with
the easy axis tilted out of the film  and in-plane projection along
$[{\bar 1}10]$ \cite{jcm}. Consequently, the Bloch domain walls are
parallel to $[{\bar 1}10]$ and the magnetic domains are in the form
of stripes. Figure \ref{pattern}a shows an image from \cite{ann} of
the stripe domain structure  with domain width of $\sim 200$ nm. The
image is obtained with transmission electron microscope in Lorentz
mode. The width of the domain wall has been theoretically estimated
to be on the order of $\sim 3$ nm \cite{ann} and recent experiments
yield consistent results \cite{asulin}.

\begin{figure}
\begin{center}
\includegraphics [scale=0.6]{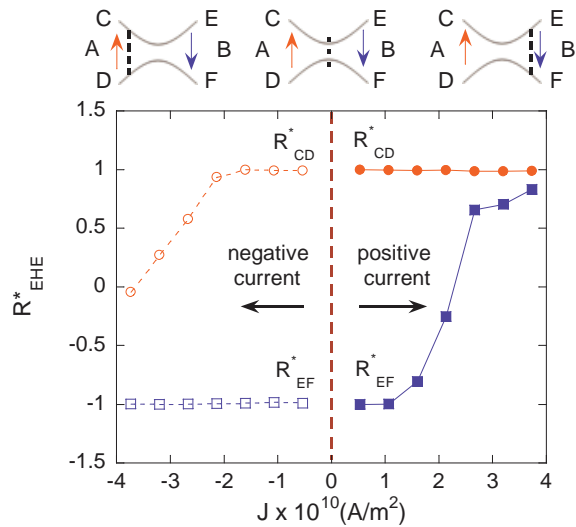}
\end{center}
\caption{Normalized EHE ($R_{EHE}^*$) measured with terminals C-D
($R_{CD}^*$) and terminals E-F ($R_{EF}^*$) as a function of
positive and negative current pulses (100 ms) at T=120 K. In the
initial state there is a single domain wall in the middle of the
narrow constriction. The right side of the graph shows the change in
$R_{CD}^*$ and $R_{EF}^*$ when pulses are applied from A to B. The
left side  of the graph starts from the same initial position and it
shows the change in $R_{CD}^*$ and $R_{EF}^*$ when pulses are
applied from B to A.} \label{currdir}
\end{figure}

The measurements presented here are of a 375 {\AA} - thick film of
${\rm SrRuO_3}$ with resistivity ratio of $\sim 20$ - indicative of
its high quality. The film was patterned using e-beam lithography
followed by ion milling.

Figure \ref{pattern}b shows the e-beam fabricated pattern whose
measurements are presented here. In our experiments we prepare a
state where only a single domain wall is between terminals A and B
and then we manipulate the position of the wall by injecting
positive and negative current pulses ($I_{AB}$) between terminals A
and B. The  magnetic state is monitored by measuring the
extraordinary Hall effect which is proportional to the average
component of the magnetization perpendicular to the film plane
\cite{EHE}. For our purposes we always present the EHE normalized to
its value when the film is fully magnetized and therefore its value
reflects the ratio between domains of opposite sign in the measured
area. We use the following notations: $R_{EHE}^*$ denotes normalized
EHE, and $R_{CD}^*$ and  $R_{EF}^*$ denote the normalized EHE
measured with terminals C-D and terminals E-F, respectively.

\begin{figure}
\begin{center}
\includegraphics [scale=0.6]{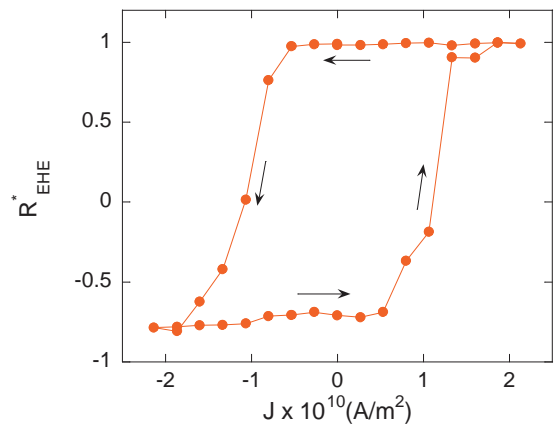}
\end{center}
\caption{Current-induced hysteresis loop with current pulses (100
ms) at T=140K and H=0 measured with normalized EHE ($R^*_{EHE}$).}
\label{RJ}
\end{figure}

In our experimental setup we deduce the location of a domain wall by
following changes in the local magnetization as determined by the
EHE measurements. Therefore, it is important to have a single domain
wall in the measured area. If there are two or more domain walls
that move in the same direction in the area monitored by a certain
pair of leads (C-D or E-F) their motion would not lead necessarily
to change in the overall average magnetization in the measured area.
Moreover, it would be impossible to deduce whether the wall moves
with or against the current.

To reach a situation with a single domain wall moving between A and
B we start by applying a large  magnetic field which fully
magnetizes the film. Then we apply an opposite field and gradually
increase its magnitude until we have nucleation of a region with
reversed magnetization. The first nucleation always occurs in the
left side of the pattern - between C and D and we increase the field
until $R^*_{CD}$ reaches the value of fully magnetized region,
namely $R^*_{CD}=1$, while $R^*_{EF}$ maintains its initial value,
namely, $R^*_{EF}=-1$. At this stage we decrease the magnitude of
the negative field and inject current pulses of 100 ms between A and
B. The magnitude of the field we leave is too small for causing wall
displacement but it serves to facilitate current-induced domain wall
movement that expands the magnetic domain from left to right. Figure
\ref{currdir} shows how this initial state is changing when current
pulses are applied from A to B (the right side of the figure) and
how the same initial stat is changing when pulses are applied from B
to A (the left side of the figure). We see that when pulses are
applied from A to B, the magnetization monitored by C-D remains
unchanged and above a current threshold a change in magnetization is observed in the right side
of the pattern, suggesting that the domain wall at the constriction
moves with the current to the right. When pulses are applied from B
to A we see that the magnetization monitored by E-F remains
unchanged and above a current threshold the magnetization monitored
by C-D starts to decrease, suggesting that the domain wall at the
constriction moves with the current to the left. The displacement
occurs at higher current densities since the current is acting
against the field.

With this experiment we show that with a single domain wall present
we can determine that domain walls move with the current and that we
can deduce  domain wall displacement from average magnetization
measurements.

A compelling demonstration of domain wall manipulation with current
is presented in Figure \ref{RJ} which shows a full current-induced
hysteresis loop with H=0. The Figure clearly demonstrates the
systematic displacement of the domain wall with current pulses
applied in opposite directions.

The hysteresis loop shows that domain wall displacement is achieved
only with current density above a certain threshold and that there
is a typical range of currents for which significant displacement is
achieved.

\begin{figure}
\begin{center}
\includegraphics [scale=0.6]{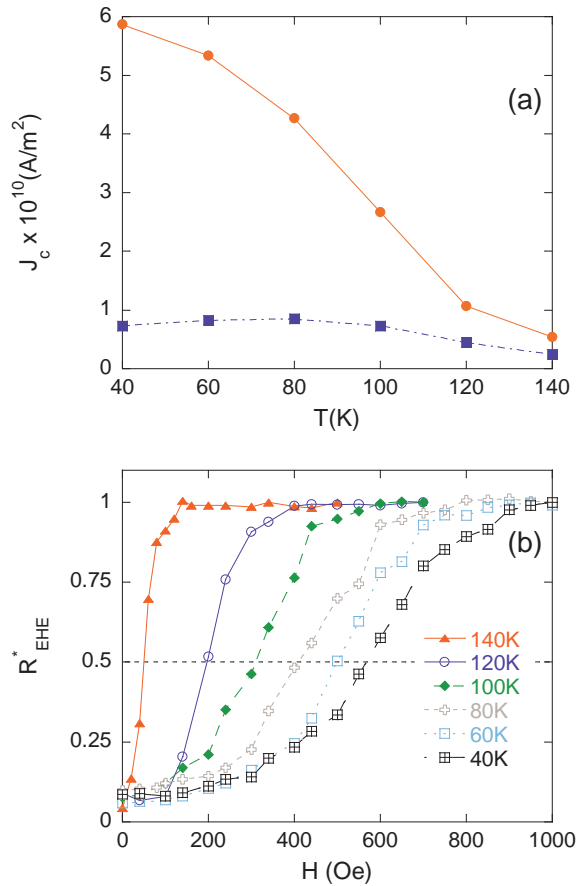}
\end{center}
\caption{(a) Characteristic measured depinning current, $J_c$, as a
function of temperature (circles) and calculated $J_c$ based on
\cite{tatara} (squares). (b) Normalized EHE as a function of applied
field. The characteristic critical depinning field, $H_c$, for each
temperature is defined as the point at which the normalized EHE is
0.5.} \label{JcHc}
\end{figure}

For quantitative study of the effect we performed similar
measurements at various temperatures. Figure \ref{JcHc}a shows
typical $J_c$ values for different temperatures. We see that $J_c$
varies between $5.3\times 10^9 \ {\rm A/m^2}$ at 140 K to $5.8\times
10^{10} \ {\rm A/m^2}$ at 40 K.

For comparing depinning currents $J_c$ at different temperatures and
in different samples it is important to exclude the effect of the
pinning potential. Since pinning potential affects depinning current
as well as depinning field $H_c$ (which is the typical field
required to move domain walls) a better comparison would be by
defining $H_c/J_c$ as a measure of efficiency in displacing domain
walls with current, since in this way the $J_c$ is normalized by the
corresponding pinning potential. For this goal we measure
field-induced hystersis loops (measured with EHE). In Figure
\ref{JcHc}b we show measurements in which the sample is cooled from
above $T_c$ at zero field to a certain temperature below $T_c$ and
the magnetization is measured as a function of the applied field. We
define the depinning field $H_c$ as the field at which the
magnetization reaches half of its fully magnetized state.

\begin{figure}
\begin{center}
\includegraphics [scale=0.6]{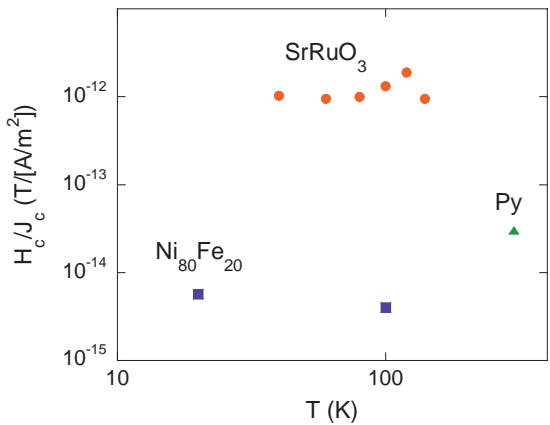}
\end{center}
\caption{$H_c/J_c$ as function of temperature in ${\rm SrRuO_3}$
(circles), Py (squares) and ${\rm Ni_{80}Fe_{20}}$ (triangle). The
data point for Py is deduced from Ref.\cite{groll} and the data
points for ${\rm Ni_{80}Fe_{20}}$ are deduced from Ref.\cite{lauf}}
\label{efficiency}
\end{figure}

Using the values of the depinning fields we calculate the efficiency
($H_c/J_c$) in  ${\rm SrRuO_3}$ for comparison with other systems
(see Figure \ref{efficiency}). We see that the efficiency in ${\rm
SrRuO_3}$ is on the order of ${\rm 10^{-12} T/[A/m^2]}$ which is
more than an order of magnitude higher than the efficiency in Py
(${\rm 10^{-13} T/[A/m^2]}$) as deduced from Ref. \cite{groll} and
in ${\rm Ni_{80}Fe_{20}}$ ring (${\rm 10^{-14} T/[A/m^2]}$) as
deduced from Ref. \cite{lauf}.

A possible reason for our results is the narrow width of domain
walls in ${\rm SrRuO_3}$. Using estimated values for the exchange
interaction and the magnetocrystalline anisotropy the width was
calculated to be on the order of ${\rm 3 \ nm}$, much smaller than
in $3d$ ferromagnetic alloys. Recently, scanning tunneling
spectroscopy measurements on thin ${\rm
SrRuO_3-YBa_2Cu_3O_7}$-bilayers   have shown
superconducting-order-parameter penetrating into ${\rm SrRuO_3}$ in
localized regions in the domain wall vicinity which puts an
experimental upper bound of ${\rm 10 \ nm}$ on the wall width. These
features make this compound a model system for studying various
properties of domain walls in the ultrathin limit.

Current-induced domain wall displacement in the limit of ultrathin
domain walls was considered by Tatara and Kohno \cite{tatara}. The
prediction is that in this limit the dominating mechanism would be
moment transfer and thus $J_c=\frac{2B\mu_B}{ena^{3}R_wA}$ where
$R_w$ is the wall resistance, $A$ is the cross-section area and
$\mu_B$ is the Bohr magneton. In our case $na^{3} \sim1$
($n\sim1.6\times10^{28}$ $[1/m^{3}]$\cite{EHE} and $a^{3}\sim
6\times10^{-29}[m^{3}]$, where n is electron density and $a$ is the
magnetic lattice constant, which is the distance between Ru ions).
Two other parameters are determined experimentally: the depinning
field $H_c$ in a procedure described above (and demonstrated in
Figure \ref{JcHc}b) and the interface resistance of the domain wall
in a procedure described by us before \cite{michael}. The calculated
values of $J_c$ according to this model are presented in Figure
\ref{JcHc}a for comparison with the experimental data. We find that
close to $T_c$ there is very good quantitative agreement
(considering the approximations used); however, in the low
temperature limit the deviation is almost an order of magnitude.
Moreover, one may have expected that when moment-transfer dominates,
the displacement would be with the electron current and against the
nominal current. Therefore, our results indicate the need to
consider other mechanisms that would be more consistent with the
experimental observations.

L.K. acknowledges support by the Israel Science Foundation founded by the Israel Academy of
Sciences and Humanities. J.W.R. grew the samples at Stanford University in the laboratory of M.R. Beasley.

\end{document}